\documentclass[12pt]{article}

\usepackage{scicite}
\usepackage{graphicx}
\usepackage{hyperref}

\usepackage{times}

\newenvironment{sciabstract}{%
\begin{quote} \bf}
{\end{quote}}

\newcounter{lastnote}

\title{From Mundane to Meaningful: AI's Influence on Work Dynamics - evidence from ChatGPT and Stack Overflow}

\author
{Quentin Gallea,$^{1\ast, 2}$\\
\\
\normalsize{$^{1}$University of Lausanne, Department of Economics}\\
\normalsize{$^{2}$Enterprise for Society}\\
\normalsize{$^{3}$EPFL, College of Management of Technology}\\
\\
\normalsize{$^\ast$ E-mail:  quentin.gallea@protonmail.com}
}

\date{}

\begin{document} 


\baselineskip24pt


\maketitle


\begin{sciabstract}

This paper illustrates how generative AI could give opportunities for big productivity gains but also opens up questions about the impact of these new powerful technologies on the way we work and share knowledge. More specifically, we explore how ChatGPT changed a fundamental aspect of coding: problem-solving. To do so, we exploit the effect of the sudden release of ChatGPT on the 30th of November 2022 on the usage of the largest online community for coders: Stack Overflow. Using quasi-experimental methods (Difference-in-Difference), we find a significant drop in the number of questions. In addition, the questions are better documented after the release of ChatGPT. Finally, we find evidence that the remaining questions are more complex. These findings suggest not only productivity gains but also a fundamental change in the way we work where routine inquiries are solved by AI allowing humans to focus on more complex tasks.
\end{sciabstract}
\newpage


\section{Introduction}
``\emph{We can only see a short distance ahead, but we can see plenty
that needs to be done.}'' Alan Turing

ChatGPT 3.5, a chatbot produced by the company OpenAI, released in
November 2022, broke the record for the fastest-growing consumer
application in history with 100 mio monthly active users in two months.
The fascination for this novel app is also accompanied by fears.
Following this release, Goldman Sachs published a report claiming that
such innovation could replace more than 300 mio jobs globally
\cite{hatzius2023potentially}.
In addition, more than 1,000 tech leaders and researchers signed a
letter calling for a pause on the most advanced AI developments.\footnote{last accessed 14.08.2023: \url{ https://www.nytimes.com/2023/03/29/technology/ai-artificial-intelligence-musk-risks.html}}

Following the citation by Alan Turing above, this article does not
intend to infer heroically on the far future of AI and its consequences but
rather delve into one of the major observable current consequences
ChatGPT has. ChatGPT is particularly good at helping us to code, from code
production to debugging. A significant amount of time if not
most of the time is spent on the internet to look for commands or
solutions to problems while coding. Debugging only is estimated to
represent about half of the time spent coding
\cite{alaboudi2021exploratory , britton2013reversible}.
Hence, any improvement in this key aspect would have important
consequences on productivity as coding is nowadays widely spread across
numerous sectors from finance to scientific research, including data science.

In this article, we explore how the release of ChatGPT 3.5 affected the
usage of the largest online coding community: Stack Overflow.\footnote{Stack Overflow is a question-answer online platform for coders.} The first
key aspect is that the release of ChatGPT was sudden, public (free
access), and occurred without the presence of any comparable model at the time. Later
other models by OpenAI (ChatGPT 4 or Code Interpreter), or by
competitors (Bard by Google) were released. Hence, focusing on ChatGPT
3.5 allows us to observe the initial shock of such models on the
worldwide coding community (not only paid users) while preventing being
confounded with the effect of other models. Despite the fact that the
release of ChatGPT was isolated from similar app releases, the effect of
ChatGPT on Stack Overflow usage could be confounded with other effects.
First, the usage of Stack Overflow could be affected by seasonality (e.g. end of the year holidays).
Second, the older the question, the higher the probability that it is
viewed and answered. Hence, views and answers are partly a function of
time.

Ideally, in order to take into account those potential remaining
confounding factors (e.g. seasonality) and measure a causal effect, we
would like to observe what would happen on Stack Overflow without
ChatGPT. While this is impossible (c.f. The fundamental problem of
causal inference), we will exploit the fact that the performance of
ChatGPT to answer coding questions is not identical for every
programming language. More precisely, we will compare the usage of Stack
Overflow between Python and R using a quasi-experimental method: Difference-in-Difference (henceforth Diff-in-Diff). Python is currently, arguably, one of the
most popular programming languages used (e.g. ranked 1st in the TIOBE
Programming Community index). Hence, it is natural to focus on this
language as it affects a large share of the coding community across
several sectors due to its versatility. In addition, it is likely that vast
resources are available to train ChatGPT to answer questions on this specific language due to its popularity. On the other hand, R, another freely accessible programming language is often compared to Python but is somewhat less versatile (initially designed for statistics) and not as widely used (e.g.~16th in the TIOBE Programming Community index). More importantly,
anecdotal evidence revealed that ChatGPT was not very efficient to
answer questions on R. Hence, R is a good potential 'control' as it is subject to seasonality or other time-varying effects on the platform while not being substantially impacted by ChatGPT.

In order to test how ChatGPT affects the way we code, we
test three hypotheses. \textbf{H1: ChatGPT decreases the number of questions asked on Stack Overflow.} People coding spend a significant
amount of time in help files and on Search Engines to look for commands
and solutions. When this strategy fails and coders are stuck, they might
turn to Stack Overflow to ask a new question. However, looking for a
solution to coding problems on an online platform takes time mainly as
one has to wait for someone else to provide the right answer. According
to an analysis of Stack Overflow in 2014, the median answer time is 16 minutes while it takes more
than a day for approximately 10\% of the questions \cite{bhat2014min}. If ChatGPT provides
answers in seconds to coding issues that we face multiple times per day, the potential economic impact could be substantial. \textbf{H2: ChatGPT increases the quality of the questions asked.} As ChatGPT brings more elements of answers,
it is likely that the remaining questions on Stack Overflow are now
better documented and researched. \textbf{H3: The remaining questions are more complex.} We can expect that the remaining questions are more challenging as ChatGPT could potentially not answer them. To test this hypothesis (H3), we are going to look at the proportion of unanswered questions as well as the average number of views per question. We conjecture that a higher proportion of unanswered questions might capture the increased complexity of the remaining questions. Additionally, we also test if the number of views per question changes. If the number of views per question remains fixed, it would provide further evidence that the increased complexity is a contributing factor to this finding, independent of the reduced platform activity.

\begin{figure}[ht!]
\centering
\includegraphics[width=0.75 \linewidth]{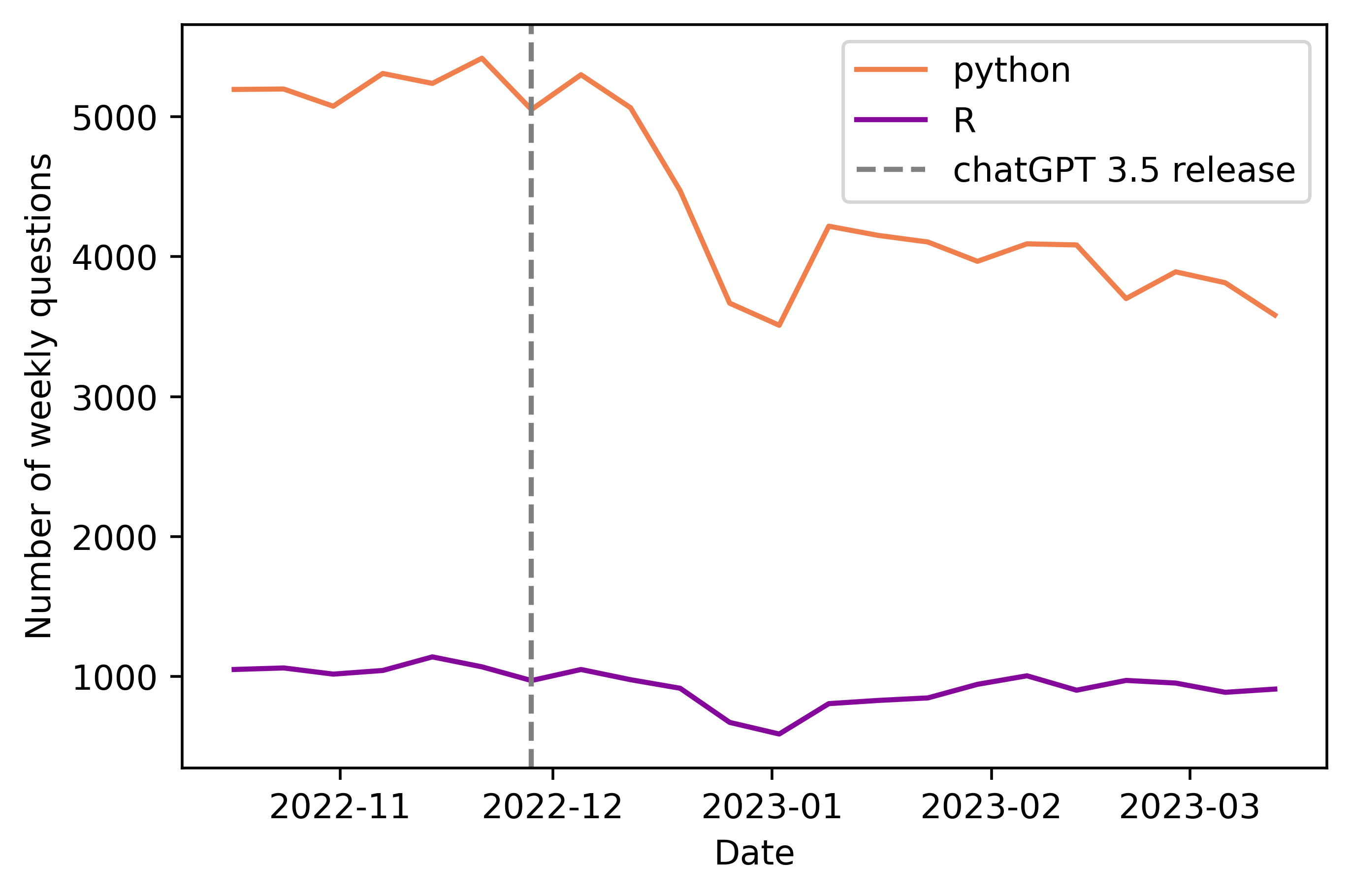}
\caption{The effect of ChatGPT on weekly number of questions on Stack
Overflow}\label{fig:fig1}
\end{figure}

Figure \ref{fig:fig1} below reveals that there is a sudden and important (21.2\%)
drop in the number of questions asked weekly on Stack Overflow about
Python after the release of ChatGPT 3.5. On the other hand, R-related questions experienced a reduction of 15.8\% during the same period. These suggestive pieces of
evidence are confirmed by the statistical model. The Diff-in-Diff model
estimates a statistically significant drop of 937.7 (95\% CI: $[-1232.8,-642.55]$ ; p-value = 0.000) weekly questions on average for Python on
Stack Overflow. Subsequent analysis using Diff-in-Diff reveals that the
quality of the questions (measured by a score on the platform) is increased and a higher proportion are left without answers. In addition, our statistical model is unable to reject the null hypothesis that there is no change in the number of views per question (p-value=0.477) and hence supports our conjecture that the complexity of the questions is increased. Hence, this paper provides evidence for the three
hypotheses defined above. These findings suggest not only productivity gains but also a shift towards more meaningful work. Indeed, by solving routine inquiries, generative AI allows humans to focus on more demanding tasks requiring expertise.

 From the industrial revolution \cite{caprettini2020rage} to the effect of AI \cite{paolillo2022compete, agrawal2019artificial, eloundou2023gpts}, including robotization at the end of the last millennium \cite{acemoglu2020robots}, technological change is known to reshape the job market significantly. Recent research revealed the effect of ChatGPT on productivity for text writing tasks \cite{noy2023experimental}. Additionally, a research paper found evidence of productivity gains on coding using GitHub\footnote{GitHub is an online platform for coders allowing to store and manage their code.} data and exploiting the ban of ChatGPT in Italy which lead to a 50\% loss of productivity two business days after the ban \cite{kreitmeir2023unintended}. An analysis of the capacity of ChatGPT for automatic bug fixing revealed that it was competitive with other state-of-the-art models \cite{sobania2023analysis}. However, another research estimated that approximately 50\% of ChatGPT coding answers had inaccuracies \cite{kabir2023answers}. Moreover, despite the promising productivity gains presented by AI, we often fail to observe those on the measure of productivity growth \cite{brynjolfsson2018artificial}. The current paper enriches the literature by highlighting the potential significant productivity improvements caused by generative AI models and how AI-Human interactions might affect the way we work. 

\section{Material and Methods}\label{material-and-methods}

\subsection{Material}\label{data}

Using the Stack Overflow Data Explorer, we obtained information on questions asked
on the platform from October 2022 to March 2023.\footnote{SQL command used to extract the data: SELECT Id, CreationDate, Score, ViewCount, AnswerCount FROM Posts WHERE (Tags LIKE '\% $ <$python$>$\% ' OR Tags LIKE '\% $<$r$>$\% ' ) AND CreationDate BETWEEN '2022-10-01' AND '2023-04-30' AND PostTypeId = 1;} The data are aggregated
at the weekly level to reduce the noise. The final sample starts on the 17th of October 2022 and ends on the 19th of March 2023.
ChatGPT 3.5 was publicly released on the 30th of November 2022. Hence,
the sample has seven weeks of data without ChatGPT 3.5 and 15 with
ChatGPT. The sample ends mid-March to avoid capturing the
effect of Google Bard released on the 21st of March 2023. Note that
ChatGPT 4 was released on the 14th of March but only for paid users.
Hence, we can expect a negligible effect of the inclusion of this last week including potential access to GPT 4 for a small subset of users (paid access).

The data includes information on the number of questions per week asked
on the platform on R and Python (see APPENDIX A for the descriptive statistics). In addition, the dataset contains the number of views, the number of answers, as well as the score for a question. The online
community allows us to measure the quality of a question by using a score. For each question, users can vote up or down.
Voting up means that you think that ``\emph{This question shows research
effort; it is useful and clear}'' or you can vote down to say that
``\emph{This question does not show any research effort; it is unclear
or not useful}''. The former increases the score by 1 point while the
latter decreases it by 1 point.

To test the last hypothesis (H3) we use the weekly proportion of
unanswered questions. Working with the proportion has the advantage that
it is not affected by the fact that the stock of questions for Python is
reduced after the release of ChatGPT.

\hypertarget{method}{%
\subsection{Method}\label{method}}

In order to address endogeneity issues, we use a
Difference-in-Difference model. The first key aspect is that we exploit
the sudden release of ChatGPT 3.5 on the 20th of November 2022. At the
time, no other similar apps were publicly available. Moreover, this
app was publicly released and freely accessible. Hence, we can observe
a global shock affecting the online coding community Stack Overflow.

Despite the exogeneity of the shock, seasonality and time could affect
the activity on the online platform as explained in the introduction and
hence could be confounded with the effect of ChatGPT release. To address this issue,
we use a Diff-in-Diff model to compare publications on R and Python. On
one hand, Python is often cited as the best substitute for R. On the
other hand, anecdotal pieces of evidence suggest that the results of ChatGPT
to answer coding questions are significantly better for Python than for
R. One reason could be that the vast amount of data available online for Python offered a richer training set for ChatGPT. 

Definition of the econometric model:

$ Y_{it} = \beta_0 + \beta_1 Python_i + \beta_2
ChatGPT_t + \beta_3 Python_i \times ChatGPT_t +
u_{it} $

where $i$ and $t$ stand respectively for the topic of the question on Stack Overflow
($i \in \{R; Python\}$) and the week. $Y_{it}$ is the outcome variable depending on the hypothesis tested: Number of questions (H1), Average question score (H2), and proportion of unanswered questions as well as the average number of views per question (H3). $Python_{i}$ is an indicator
variable taking the value 1 if the question is related to Python and 0
otherwise (related to R). $ChatGPT_{t}$ is an indicator variable
taking the value 1 from the release of ChatGPT and onwards and 0
otherwise. $u_{it}$ is an error term clustered at the coding language
level ($i$).

The key identifying assumption to measure a causal effect in the
Diff-in-Diff model is called the parallel trends. This assumption means
that without treatment (here ChatGPT release), the trends between Python
and R would be parallel. As this is impossible to observe (c.f. the fundamental
problem of causal inference), we can only test if before the release the
trends were parallel. Placebo tests on the pre-ChatGPT period could not
reject the parallel trends assumption as the p-values of the two coefficients for two different tests are respectively 0.722 and 0.397 (see APPENDIX B.).

\section{Results}\label{results}

\subsection{H1: ChatGPT decreases the number of questions asked on Stack Overflow}

\begin{figure}[ht!]
\centering
\includegraphics[width=0.75 \linewidth]{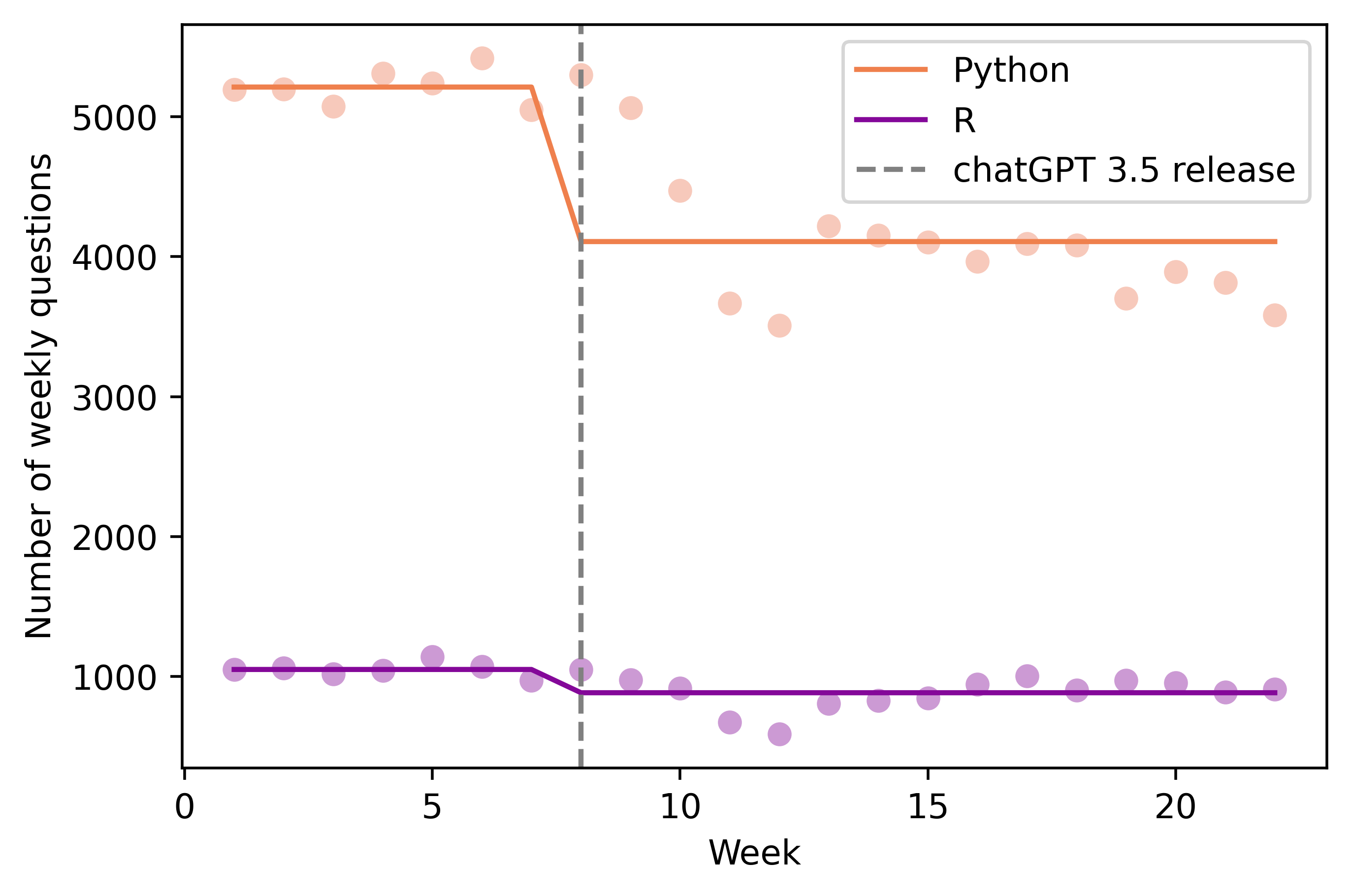}
\caption{The effect of ChatGPT on the weekly number of questions}\label{fig:fig2}
\end{figure}

Using the Diff-in-Diff model we find a statistically significant drop of 937.7 (95\% CI: [-1232.8,	-642.6 ] ; p-value = 0.000) weekly questions on average for Python on Stack Overflow (see Figure \ref{fig:fig2}). With an average number of weekly questions of 5220.6 for the period before the release of ChatGPT, this represents a reduction of 18\% of weekly questions (937.7/5212.0=0.180).\footnote{The percentage cited in the introduction of 21.2\% is the raw reduction percentage using only Python data and not Diff-in-Diff estimation.}

\subsection{H2: ChatGPT increases the quality of the questions asked}

\begin{figure}[ht!]
\centering
\includegraphics[width=0.75 \linewidth]{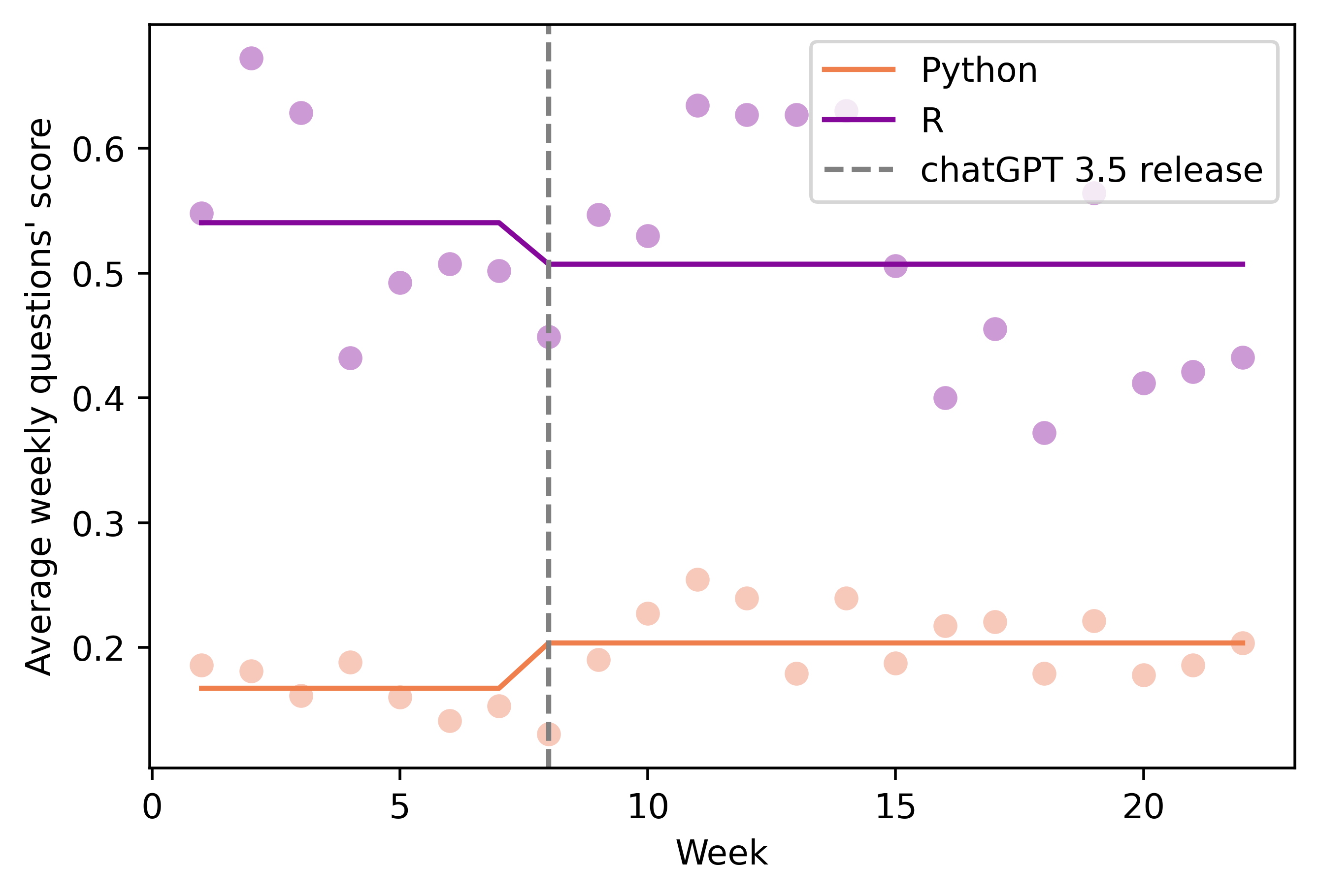}
\caption{The effect of ChatGPT on quality of the questions}\label{fig:fig3}
\end{figure}

While the overall number of questions asked drops, it is likely that the
nature of the remaining questions change. In particular, we could expect
that more basic questions could be answered quickly with ChatGPT or
ChatGPT could bring some element to answer questions without solving it
completely. If this is the case, the quality of the questions asked on
Stack Overflow should be affected positively. The second Diff-in-Diff
regression estimates that there is a 0.07 points (95\% CI: {[} -0.0127 , 0.1518 {]}; p-value: 0.095) increase of the questions' score on average (see Figure \ref{fig:fig3}). A
higher score indicates that \emph{``This question shows research effort;
it is useful and clear''} (see Section Data for more details). The
pre-ChatGPT average score per question for Python was 0.17. Hence, this
effect represents a 41.2\% increase (0.07/0.17=0.412).

  \subsection{H3: The remaining questions are more complex}

\begin{figure}[ht!]
\centering
\includegraphics[width=0.75 \linewidth]{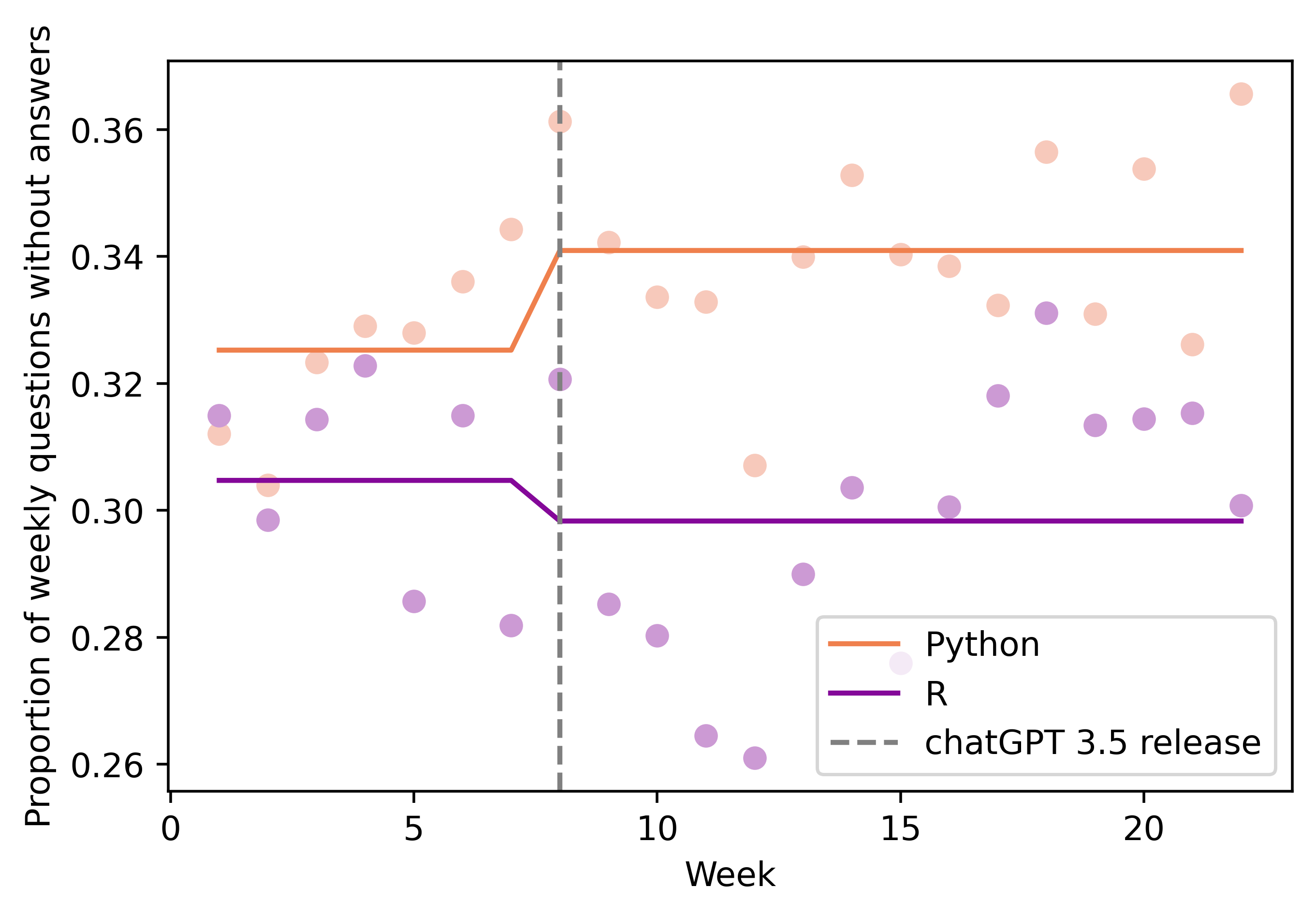}
\caption{The effect of ChatGPT on the proportion of unanswered question}\label{fig:fig4}
\end{figure}

If ChatGPT is able to solve numerous questions it is expected that the
remaining ones are the more complex. Hence, the final inquiry looks into
the change in the proportion of questions left unanswered (suggesting
that they are more complex). Recall that our focus on the proportion is designed to prevent the confounding influence of the reduced platform activity, which could potentially impact the number of answers. The Diff-in-Diff regression estimates that
there is a 2.21 percentage points (95\% CI: {[} 1.2, 3.0 {]}; p-value:
0.039) increase in the proportion of questions unanswered (see Figure \ref{fig:fig4}). The
pre-ChatGPT proportion of questions unanswered in the sample for Python
was 32.5\%. Hence, this effect represents a 6.8\% increase
(2.21/32.5=0.068).

\begin{figure}[ht!]
\centering
\includegraphics[width=0.75 \linewidth]{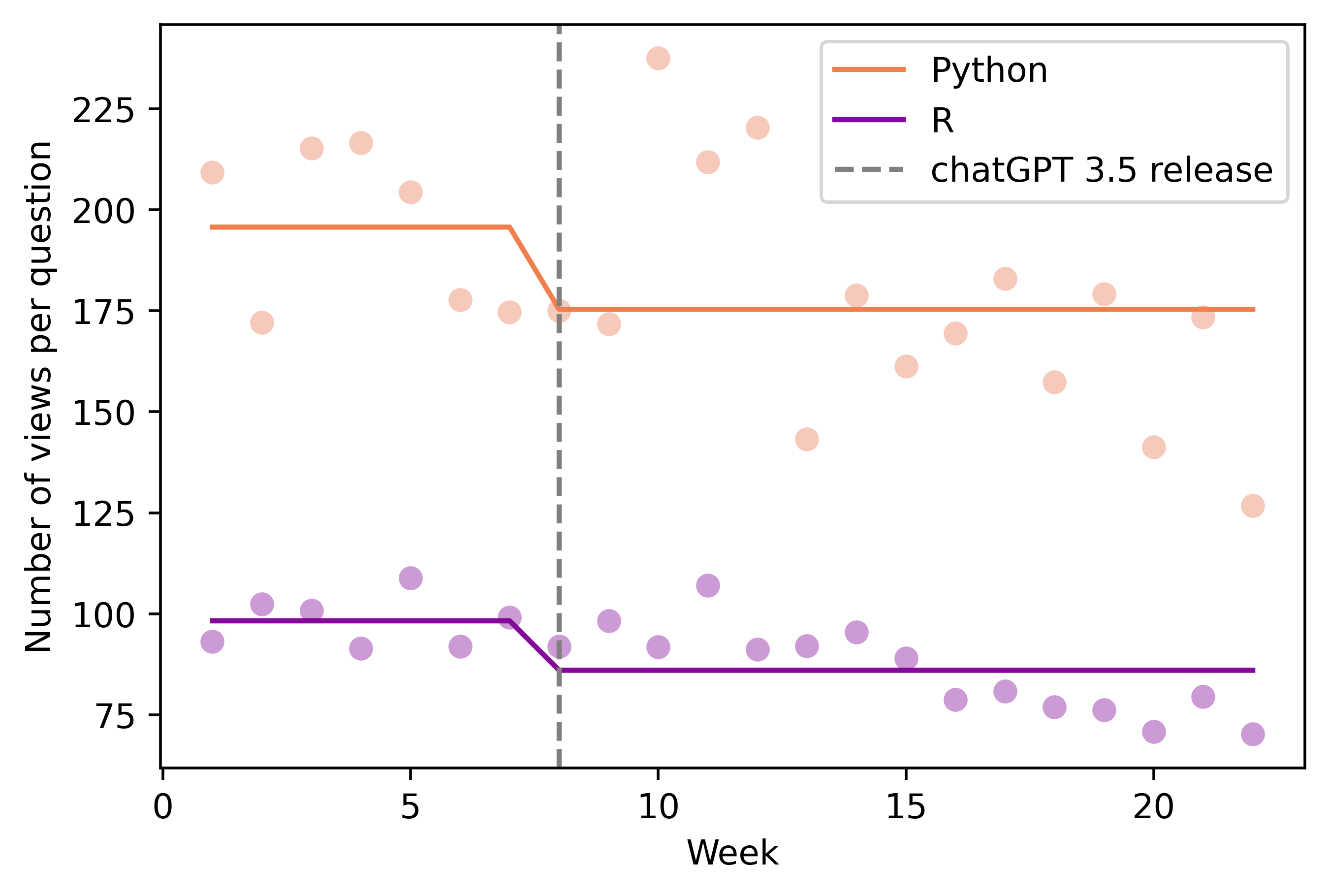}
\caption{The effect of ChatGPT on the number of views per question}\label{fig:fig5}
\end{figure}

If the number of views is falling faster than the number of questions, the higher proportion of unanswered questions could be caused by the reduced activity on the online platform. This alternative explanation would question the interpretation that it is a consequence of the increased complexity. However, if the number of views per question is not changing after the release of ChatGPT, it would support our interpretation of H3. Using our Diff-in-Diff model to test if there is a change in the average number of views per question confirms our intuition by revealing no statistical difference after the release of ChatGPT (coefficient= -8.126; 95\% CI: {[} -31.01, 14.2{]}; p-value: 0.477; see Figure \ref{fig:fig5}).

\clearpage
\section{Discussion}\label{discussion}

The rapid evolution of generative AI technologies is reshaping the way we approach numerous tasks, in particular, offering new tools that have the
potential to revolutionize coding practices and enhance productivity.
Our investigation focused on the impact of ChatGPT 3.5, an advanced AI
chatbot released by OpenAI, on the dynamics of coding problem-solving
within the context of the largest online developer community, Stack
Overflow.

This paper reveals that ChatGPT 3.5's release in November 2022 resulted in a substantial
reduction of the number of questions on Stack Overflow particularly
those related to the Python programming language (estimated to be about one-fifth in
this research). Concurrently, we observed an increase in the quality of
questions that remained on the platform, accompanied by a rise in the
number of unanswered questions while the average number of views per question remained stable. 

\subsection{Limitations}

The current analysis does not dive into the nature of the users which opens several questions. First, we do not know if the reduction in the number of questions asked online concerns any profile or more or less skilled workers. Recent findings on the effect of ChatGPT on text writing tasks revealed that low-skilled workers benefited the most from such tools \cite{noy2023experimental}. Secondly, if routine tasks are solved by AI, will it boost the efficiency of lower-skilled jobs or will it replace them? Previous research findings favored the former \cite{autor2015there, bessen2019automation}. However, given the novelty of such technology, it would be safer to reassess this particular case. Third, does it helps particularly in the initial learning phase, rather than while practicing or in both situations? Again, this consideration is important to establish who is benefiting the most from such tools and for what usage. Hence, once we have a deeper understanding of who benefits and how, a quantification of productivity gains could be made.

\subsection{Implications}

The reduction in the volume of questions with an increase in the quality and potentially of the complexity of the remaining questions on Stack Overflow raises important implications for time management and resource allocation. With ChatGPT efficiently addressing a significant portion of `basic' inquiries, people coding can focus on more complex challenges that require human expertise. While our study focuses primarily on the impact on the Python language, suggests a broader paradigm change, where AI tools become instrumental in managing routine tasks and enable humans to focus on higher-value tasks.

While the immediacy of ChatGPT is likely to improve productivity in the short run, this efficiency could affect negatively how we learn and tackle problems by solving for us most of the challenges. Finally, if Stack Overflow is an important source for the training sample of ChatGPT and alike, the stark reduction in content might affect the long-term performance of such models. Hence, the question of the long-term net effect on society remains open and future research should focus on solving this puzzle.


\newpage

\bibliography{scibib}

\bibliographystyle{Science}
\newpage

\section*{APPENDIX}

\subsection*{A. Descriptive statistics}

\begin{table}[ht!]
\caption{Weekly average by period and programming language\label{tab1}}
\centering
\vspace{4ex}
\begin{tabular}{c | l | cc}
\multicolumn{1}{l}{}    &           & \multicolumn{2}{c}{\textbf{ChatGPT release}} \\
\multicolumn{1}{l}{}    &           & \textbf{pre}          & \textbf{post}         \\
\hline
\textbf{Python} \rule{0pt}{4ex}   & Questions & 5212.000      & 4108.133      \\
                         &                      & (127.604)     & (509.034)    \\
                   \rule{0pt}{4ex}       & Score      &  0.167        & 0.204        \\
                                          &           & (0.018)      & (0.032)      \\
                   \rule{0pt}{4ex}       & No answer  & 0.325        & 0.341        \\
                                         &           & (0.014)      & (0.015)      \\
                        \hline
\textbf{R}    \rule{0pt}{4ex}     & Questions       & 1051.143      & 884.933       \\
                          &                          & (51.577)      & (122.891)     \\
                \rule{0pt}{4ex}          & Score      & 0.540        & 0.507        \\
                          &                           & (0.084)      & (0.094)      \\
                 \rule{0pt}{4ex}         & No answer  & 0.305        & 0.298        \\
                          &                             & (0.016)      & (0.021)     \\
                          \hline \hline
\multicolumn{4}{l}{Standard deviations in parenthesis.}
\end{tabular}
\end{table}

\subsection*{B. Parallel trends assumption}

In order to test if the trends are parallel in the pre-ChatGPT period, we run two placebo tests. The sample is restricted to the pre-ChatGPT period. Then, we define set an indicator variable equal to one from week three and onward as well as a second one on week four and onward. Finally, we run our Diff-in-Diff model on this sub-sample using as a treatment the two different placebo periods. Based on the results of the regressions, it is not possible to reject the parallel trends (the p-values of the two coefficients are respectively 0.722 and 0.397).

\begin{table}[ht!]
\caption{Placebo tests\label{tab2}}
\centering
\begin{tabular}{lcc}
                  \rule{0pt}{4ex}             & \textbf{(1)} & \textbf{(2)}  \\
\hline
\textbf{Dep. Variable}         &   question   &   question    \\
\textbf{Estimator}             &   PanelOLS   &   PanelOLS    \\
\textbf{No. Observations}      &      14      &      14       \\
\textbf{Cov. Est.}             &  Clustered   &  Clustered    \\
\textbf{R-squared}             &    0.9981    &    0.9984     \\
\textbf{R-Squared (Within)}    &    0.0066    &    0.1474     \\
\textbf{R-Squared (Between)}   &    1.0000    &    1.0000     \\
\textbf{R-Squared (Overall)}   &    0.9981    &    0.9984     \\
\textbf{F-statistic}           &    1789.0    &    2085.0     \\
\textbf{P-value (F-stat)}      &    0.0000    &    0.0000     \\
\hline
\textbf{Intercept}             &    1056.5    &    1043.7     \\
\textbf{ }                     &   (229.59)   &   (81.715)    \\
\textbf{placebo1}              &   -7.5000    &               \\
\textbf{ }                     &  (-0.2490)   &               \\
\textbf{python}                &    4140.0    &    4112.3     \\
\textbf{ }                     &   (867.98)   &   (99.894)    \\
\textbf{placebo1:python}       &    29.200    &               \\
\textbf{ }                     &   (0.3667)   &               \\
\textbf{placebo2}              &              &    13.083     \\
\textbf{ }                     &              &   (0.3445)    \\
\textbf{placebo2:python}       &              &    84.917     \\
\textbf{ }                     &              &   (0.8843)    \\
\hline \hline
\multicolumn{3}{l}{T-stats reported in parentheses}
\end{tabular}
\end{table}

\end{document}